\documentclass[aps,prl,floatfix,superscriptaddress,twocolumn,showpacs]{revtex4}

\usepackage{epsfig}
\usepackage{graphicx,psfrag}
\usepackage{dcolumn}
\usepackage{bm}
\usepackage{amssymb}
\newcommand{\nn}{\nonumber \\}

\def\L{\mathcal{L}}

\def\d{\partial}

\def\le{\left}
\def\ri{\right}

\def\be{\begin{equation}}
\def\ee{\end{equation}}
\def\bei{\begin{itemize}}
\def\eei{\end{itemize}}
\def\beq{\begin{eqnarray}}
\def\eeq{\end{eqnarray}}

\def\tu{\tilde{u}}
\def\tv{\tilde{V}}

\begin{document}
\title{Generating the mass gap of the sine-Gordon model}

\author{V. Pangon}
\affiliation{Gesellschaft f\"ur Schwerionenforschung mbH, Planckstr. 1, D-64291 
Darmstadt, Germany}
\affiliation{Frankfurt Institute for Advanced Studies, Universit\"at Frankfurt, 
D-60438 Frankfurt am Main,
Germany}
\affiliation{{\it on leave to :} DiscInNet Labs, 5 rue de l'Eglise,92100 Boulogne, France}

\begin{abstract} 
We discuss in this study the possibility of finding a finite mass gap in the broken phase  of the sine-Gordon model in $d=2$ using the functional flows. 
We demonstrate that the signal of the presence of massive excitations, a finite positively-curved {\em blocked} potential around its minima, is recovered only in our treatment. The usual results based on the flow of the Fourier expansion of the {\em blocked} action are then shown to actually fit a singularity.
\end{abstract}
\pacs{11.10.Gh,11.30.Qc,11.10.Kk,05.10.Cc}

\maketitle
{\bf I. Introduction}
- The sine-Gordon Model (sGM) whose bare Lagrangian in $d$ euclidean dimensions reads :
\be\label{1}
\L=\frac{z_B}{2}\d_\mu\phi\d_\mu\phi+u_B\,cos(\phi)
\ee
has attracted a considerable interest these last decades due to its unique features. In $d=2$, the Coleman point $(u_c,z_c)=(0,\frac{1}{8\pi})$ plays a special role in the weak-coupling $u_B$ limit and provides a three-sectors  phase diagram : a massless sector for $z_B<z_c$ due to a vanishing interaction, and a strong-coupling one for $z_B>z_c$. These two sectors are separated by a cross-over regime around $z_B\simeq z_c$ for finite $u_B$ and the model is known to undergo a Kosterlitz-Thouless phase transition  \cite{Kosterlitz}. The spontaneously broken periodicity symmetry phase corresponds to the two last sectors and gives rise to dynamical mass generation.  It  exhibits  solitons and bound-states and is  quantum S-dual to the massive Thirring model  where its solitons correspond to the fermions of the latter \cite{Coleman}. The exact spectrum of these topological excitations is known \cite{Arefeva}. 
At the partition function level, it is also dual to the classical neutral Coulomb gas in all dimensions and  is thus of great interest to study the roughening transition   \cite{Samuel} and the universal jump   of the superfluid density in $d=2$. Finally, the sGM is also used to describe the thermal fluctuations of membranes in a periodic pinning potential (e. g. \cite{nozieres}).

Because the sGM has been exactly solved, it is an ideal testing ground to understand better theoretical approaches, such as functional Renormalization Group (fRG). Understanding how fRG works for the sGM  will provide us new insights to study models where the exact methods are beyond our abilities but fRG is immediatly applicable, such as a more general periodic interaction an/or in another dimension.\\
The present Letter focuses on the signature of the dynamical mass generation, which is the central phenomenom occuring in the IR of the broken phase. We will demonstrate that the appearance of this new mass scale can be seen only in our approach.

{\bf II. Functional flows}
- In the following, we use the Effective Average Action (EAA) method.  Its strategy is to build a continuous set of functionals interpolating smoothly between the bare action of the theory one whishes to solve and its effective action. These functionals $\Gamma_k$ --the effective average action-- are demonstrated to follow the flow equation : 
\be
k\d_k\Gamma_k=\frac{1}{2}Tr\le( \frac{k\d_k R_k(p)}{\Gamma^{(2)}(p)+R_k(p^2)}\ri)
\ee
where the regulator $R_k(p^2)$ is a well-chosen function of $k$ and $p^2$ \cite{Wetterich:1992yh}.
Obviously, the functional equation (2) is in general too difficult to be solved directly, and calls for further assumptions.
The most common approximation is to project the EAA on the functional ansatz :
\be
\Gamma_k[\phi]=\int d^2x\frac{Z_k[\phi]}{2}\d^\mu\phi\d^\mu\phi+V_k[\phi]
\ee which gives now a set of two coupled flow equations for $V_k(\phi)$ and $Z_k(\phi)$  (see e. g. \cite{Seide}). We solved this system with no further approximation for the first time using the algorithm \cite{BerzinsDew2} for the power-law regulator $R_k(p^2)=p^2\le(\frac{p^2}{k^2}\ri)^{-b}$ with $b\geq1$ and we discuss in the following   how to compare these results with the usual perturbative approach and how to extract new physical informations.
\\{\it a. The perturbative flow.} 
Instead of solving directly the flow of $V_k(\phi)$ and $Z_k(\phi)$, the perturbative approach projects their evolution onto their first harmonic in the small $\tu_k$ limit \cite{Nagy} :
\beq
k\d_k \tu_k&=&\frac{1}{\pi} \int_{-\pi}^{+\pi}\!\!\!\! \!\!\!\!d\phi\, k\d_k \tv_k(\phi)cos(\phi)\nn
&=&-2\tu_k+\le(\frac{\tu_k}{z_k}\ri)\sum_{n=0}^\infty A_{2n+1}(b)\le(\frac{\tu_k}{z_k^{1-1/b}}\ri)^{2n}\\
k\d_k z_k  &=&\frac{1}{2\pi} \int_{-\pi}^{+\pi}\!\!\!\!\!\!\!\! d\phi\, k\d_k Z_k(\phi)\nn
&=&\sum_{n=1}^\infty B_{2n}(b)\le(\frac{\tu_k}{z_k^{1-1/b}}\ri)^{2n}
\eeq where the tilde stands for dimensionless quantities.
The weak-coupling expansion is thus actually not driven by $\tu$ but rather by $\frac{\tu}{z^{1-1/b}}$.
The coefficients can be systematically computed analytically, and two  universal --regulator independent-- features appear. The Coleman point is always reproduced as demonstrated in \cite{Pangon:2010uf} i.e.  $A_1=\frac{1}{4\pi}$  and the coefficient $B_2$ is negative, providing a positive anomalous dimension.
The set of differential equations (4)-(5) yields an RG invariant $\tau$ that can be explicitely computed at order $\tu_k^2$ (and $\tu_k^3$) e. g. for $b=2$ :
\be
\tau=\tu_k^2- \frac{6}{z_c}\le(\frac{z_k}{z_c}-1\ri)^2
\ee
 and thus provides analytically the trajectories in the $(\tu_k,z_k)$-plane of the sytem.
These trajectories can be casted in three categories, corresponding to the three sectors  : 
the massless phase for $\tau<0$ and $z_k<z_c$, the strong-coupling one for $\tau<0$ and $z_k>z_c$ and the cross-over regime for $\tau>0$. The latter is limited by the separatrix $\tau=0$, plotted on fig. 1. On the same figure, we draw a perturbative trajectory of the strong-coupling sector and for comparison a serie of trajectories coming from our full solution of $V_k(\phi)$ and $Z_k(\phi)$ defining the couplings $\tu_k=\frac{1}{\pi} \int_{-\pi}^{+\pi}\!\!\!\!d\phi\,  \tv_k(\phi)cos(\phi)$ and $z_k=\frac{1}{2\pi} \int_{-\pi}^{+\pi}\!\!\!\! d\phi\,  Z_k(\phi)$. 
\begin{figure}[h!]
\begin{center}
\includegraphics[angle=270,scale=0.50]{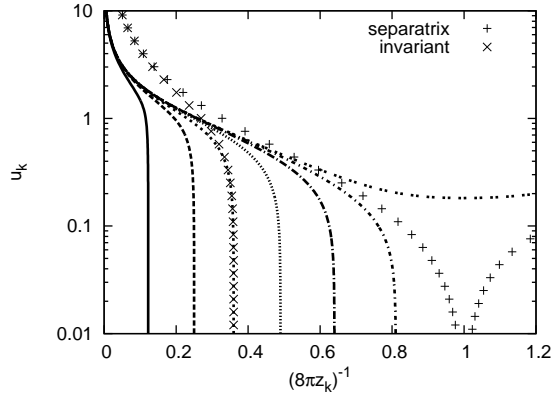}
\caption{Trajectories from the full solution of (3) projected onto the $(\tu_k,z_k)$ plane in the strong-coupling and the cross-over sectors (lines), the separatrix defined by $\tau=0$ ($+$) and a perturbative trajectory ($\times$) for $b=2$.}
\end{center}
\end{figure}

 The UV flow of the full solution and the perturbative result are in good agreement and  one can thus use the RG invariant (6)  to extract the correlation length of the dual XY model, and the vortex-vortex two-points function critical exponent in the vicinity of the Coleman point \cite{Nagy}.  On the contrary, the IR flow is not driving the system in the close neighborhood of the separatrix, as perturbatively expected. As a result, the use of the RG invariant $\tau$ to give a rough estimate of the soliton mass \cite{Amit} is incorrect.
\\{\it b. Breakdown of the perturbative approach.}
 The discrepancy between  the perturbative results and the full solution  has many origins. We made a small $\frac{\tu_k}{z^{1-1/b}}$ expansion, which turns out to be large in the IR even if  both couplings $\tu_k$ and $z_k$ are growing as $k$ is decreased, due to the $\beta$-function of $\tu_k$ winning the competition with the one of $z_k$.
As a result, truncating the equation (4)-(5) at the second (or any finite) order turns out to be  illegal in the IR and one has to solve the first lines of (4) and (5) without expansion \cite{Nagy}.
In addition,  taking into account only the first harmonic of period $2\pi$ of the potential $V_k$ uses the relevance classification in the UV, neglecting the appearance of higher period harmonics. This is certainly legal at the beginning of the flow, but it is less and less correct as the non-linearities generate these harmonics and as $z_k$ grows :  below the scale $k$ for which $z_k=n^2z_c$, the terms in $\propto cos(n\phi)$ in $V_k$ will start being relevant.
Beyond this technical issue is actually hidden the fact that the IR fluctuations are expected to be dominated by the tunneling between different equivalent minima --this is the symmetry broken phase-- i. e. the contribution of the topological excitations. Dropping terms of period $2n\pi$ neglects a part of the contributions that the topological configurations of winding number $|n|$ should see. As a result, using a finite order Fourier expansion of $V_k(\phi)$ is dangerous. We now demonstrate that it is also the case for $Z_k(\phi)$ if one wants to see the effect of the mass of the excitations.

{\bf III. Signature of the excitations} - 
As it has been emphasized in \cite{Pangon:2010uf} already at Local Potential Approximation (LPA), the behaviour of the dimensionful quantities such as the coupling $u_k$ are not informative at all in the case of the sGM, because the effective potential --that corresponds to $V_{k=0}(\phi)$-- is a constant in both phases, being the only possibility for a periodic and convex function of its argument. This tells us that what counts is how the effective potential is reached i. e. the difference is made on the dimensionless  potential $\tv_k$. 
\\The minima of the bare potential (1) are located in $v=\pi+2n\pi$ and we expect that the quantum vacua, solutions of the equation of motion for the effective action are sitting at the same location. The vacua are in our case actually built progressively from the minima of $\Gamma_k$, also located in $v$. In the broken phase, for $k$ large enough, we feel the contributions of the topological excitations which tend to increase slowly $\tv_k''(v)$  with $k$ (minima selection) but their mass are roughly priceless in units of $k$. For scales $k$ lower than say $\frac{1}{\xi}$, we start feeling the mass of some of the excitations and any attempt of reaching an excited state has thus a finite cost in $k$, accelerating the flow of $\tv''_k(v)$ (massive cross-over).
 Finally, for $k<<\xi^{-1}$, one fully feels the mass of the excitations and the energy needed to feed excited states is constant i. e. $\tv''(v)$ diverges as $k^{-2}$ as $k$ is descreased. \\
{\it a. The Fourier expansion of $V_k$ and $Z_k$.}
 As a result, one realizes that the signature of the excitations can be seen only if one can carefully study the close neighborhood of $v$ of $\Gamma_k$. Because the bare theory (1) is periodic and this periodicity is preserved by (2) and (3), it would seem natural to represent $V_k$ and $Z_k$ with their Fourier series. The problem is that already at LPA, this strategy is known to fail for $V_k(\phi)$ \cite{Pangon:2010uf}. The reason is that the expansion does not converge due to the severe competition between the convex regions --i.e. around the minima, where the potential is well-behaved-- and the concave ones --around the extrema, where the potential lives near a singularity, see below. This abrupt change of behavior between the convex and the concave regions for the potential $V_k(\phi)$ is thus expected to propagate to $Z_k(\phi)$ when we solve (3) that couples the two evolutions.\\
\begin{figure}[h!]
\begin{center}
\includegraphics[angle=270,scale=0.49]{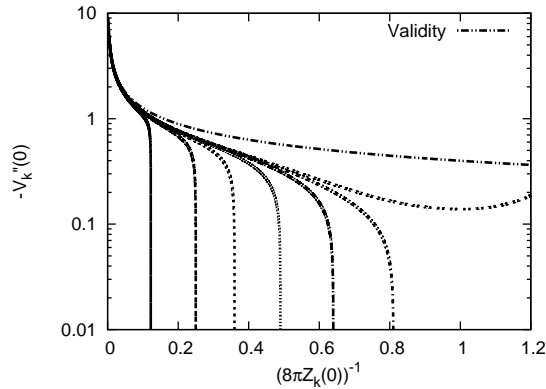}
\caption{The same trajectories as in fig. 2 projected  onto the $(-\tv''_k(0),Z_k(0))$ plane for $b=2$. The extra trajectory traces the limit of validity of the approach given by $\alpha=0$, see (7).}
\end{center}
\end{figure}
{\it b. Instability driven Fourier flow.}  Because the Fourier expansion is probably breaking down, it is informative to compare the value of $V_k(\phi)$ and $Z_k(\phi)$ from our full solution at various $\phi$ and the value of the Fourier coefficients $\tu_k$ and $z_k$ one can build from them. At the special point $\phi=0$ that places us in the middle of the concave regions, one can plot the trajectories in the $(-\tv''(0),Z_k(0))$ plane \footnote{At leading order in the Fourier expansion it coincides with $\tu_k,z_k$} on fig 2 and realize that it has precisely the same shape as the usual phase diagram of fig. 1. The IR flow now gets close to the line of equation
\be
\alpha=1+\tv''_k(0)Z_k(0)^{-1+1/b}\le(b-1\ri)^{1-1/b}b^{-1}\gtrsim0
\ee which is the limit of validity $\Gamma_k^{(2)}+R_k=0$ for (2) under the assumption (3) when using our regulator. 
\begin{figure}[h!]
\begin{center}
\includegraphics[angle=270,scale=0.50]{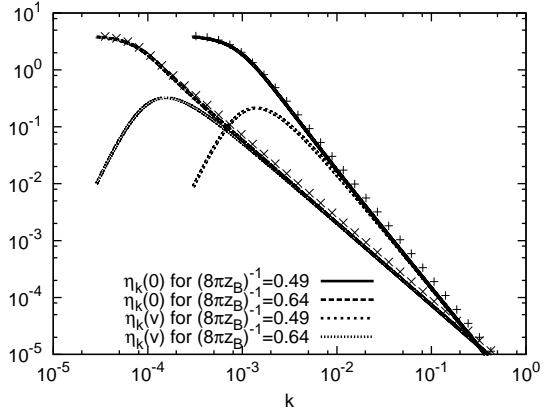}
\caption{The running of the anomalous dimensions $\eta_k(\phi)$ in $\phi=0$ and $\phi=v$ (lines) together with the Fourier reconstructed values (crosses) as a function of $k$ for $b=2$ .}
\end{center}
\end{figure}
Therefore, the Fourier expansion of $\Gamma_k$, even when valid, is actually not only mimicking the concave region, but also fitting a singularity and thus is not physically relevant. A simple explanation is provided by the anomalous dimension
\be
\eta_k(\phi)=-k\d_k Log(Z_k(\phi))
\ee computed for $\phi=0$ and $\phi=v$, and compare them with the one from the reconstructed Fourier coefficient $z_k$. The figure 3 shows that in the UV, the behavior is the same in convex and concav region and in good agreement with the Fourier representation, but dramatically differ in the IR. In particular, it shows that the anomalous dimension goes to 0 around the minima while it tends to a large value in the concave regions. This large value in the concave regions makes them count more and more when going IR, which explains why they are the ones seen by the Fourier expansion. It also explains why even if one solves fully $V_k(\phi)$ together with $z_k$ defined as in the first line of (5), one can not stabilize the flow around the minima : because $z_k$ -- that fits the concave regions behavior-- keeps running in the flow equation for $V_k$, as in \cite{nozieres,Nagy}.
\begin{figure}[h!]
\begin{center}
\includegraphics[angle=270,scale=0.50]{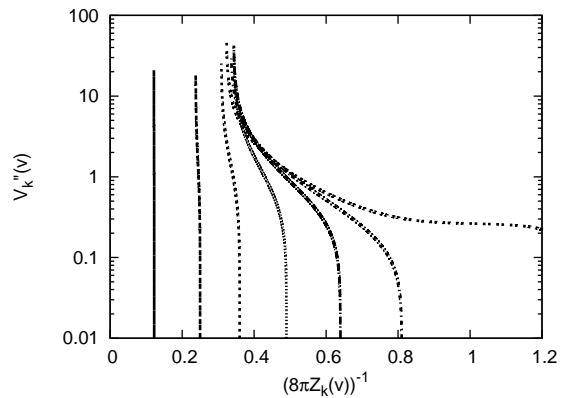}
\caption{The physical phase diagram : the same trajectories as in fig. 1 and 2 projected  in the $(\tv''_k(v),Z_k(v))$ plane for $b=2$. The focalisation effect in $(8\pi Z_k(v))^{-1}\simeq 0.35$ is not genuine, as one can pick different bare theories converging to  other IR values.}
\end{center}
\end{figure}

{\bf IV. The physics around the minima} - We can finally turn our attention to see how $\Gamma_k$ behaves near $v$ to get the physical phase diagram that is plotted  in the $(\tv''_k(v),Z_k(v))$-plane [14]. The figure 4 exhibits actually a similar behaviour in a much more flexible approach as the one obtained in a different context, using Wegner's blocking \cite{Kehrein}. As expected, the values of $Z_k(v)$ saturates in the IR, in agreement with fig. 3.
\begin{figure}[h!]
\begin{center}
\includegraphics[angle=270,scale=0.50]{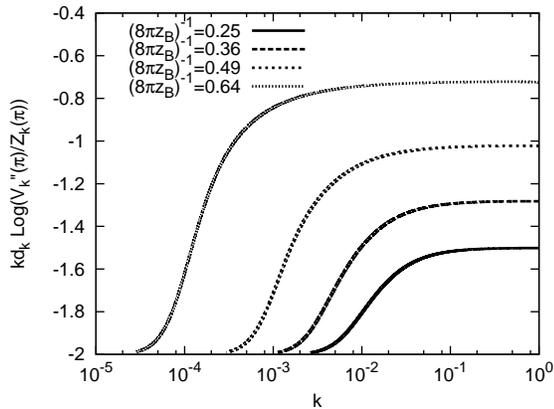}
\caption{The flow of the quantity $k\d_k Log\le(\frac{\tv''_k(v)}{Z_k(v)}\ri)$ for $b=2$ as a function of $k$. It saturates in the IR to the value $-2$, showing that $\frac{\tv''_k(v)}{Z_k(v)}\propto \frac{1}{k^2}$}
\end{center}
\end{figure}
The proof of mass generation in the IR is given by the saturation of the curvature $V_k''(v)$ provided that $Z_k(v)$ stops running. A natural quantity to probe this is $k\d_k Log\le(\frac{\tv''_k(v)}{Z_k(v)}\ri)$ which should go to $-2$ in the case of a finite dimensionful curvature. As expected, the figure 5 shows that we have been able to stabilize a finite mass gap never seen before in other renormalization procedures. As we saw, its obtention heavily relies not only on truncating of the flow equation (2) non-perturbatively in a consistent manner (3) to access the strong-coupling regime, but also on not using the relevance classification that is UV i.e. the Fourier expansion of $\Gamma_k$. It illustrates nicely the fact that the relevance classification can be totally changed in the IR even for a massive theory. We also emphasize that the topological excitations responsible for the mass gap were never explicitly encoded in this study, showing the power of fRG even away from criticality of a Kosterlitz-Thouless transition \cite{Gersdorff}. 
\begin{figure}[h!]
\begin{center}
\includegraphics[angle=270,scale=0.50]{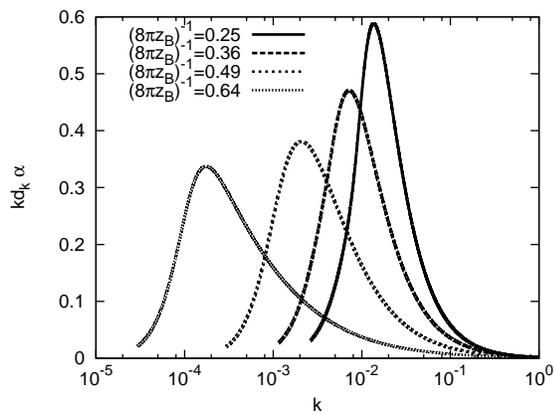}
\caption{The $\beta$-function of $\alpha$ as defined in (7) as a function of the scale $k$.}
\end{center}
\end{figure}
\\We conclude with the problem of the stability of the IR flow. We saw on fig. 2  that the flow in the concave regions lives near the limit (7). We have not been able to stabilize the trajectory as far as one wants in the IR as in \cite{Pangon:2010uf} but one can compute the $\beta$-function of $\alpha$ given by (7) which corresponds to the evolution of the distance of the system to the instability. We found that this $\beta$-function is actually suppressed when the mass gap is built, suggesting the stabilization above the unstable line $\alpha=0$, see fig. 6. 
Intuitively, this probable IR stability can be understood quite simply. In the IR, the regulator mostly behaves as a mass term so that the EEA actually solves a serie of massive sGM whose mass parameter is progressively decreased. As it is well known, the massive sGM has the same structure of flows as the sGM for scales above its mass so that the interpolation between these models is smooth, corresponding to a cross-over located at lower and lower energies.

All the results presented here are stable against a change of regulator (changing $b$ or taking an exponential suppression) or the use of the propertime scheme \cite{Mazza:2001bp}

\begin{acknowledgments}
The author wishes to thank J.-P. Blaizot, B. Delamotte and J. Polonyi for many fruitful discussions.
\end{acknowledgments}

%
%
\end{document}